# Credit Expansion, Land Speculation, and Low-Interest-Rate Policy

by

Tomohiro Hirano and Joseph E. Stiglitz

**Abstract**

This paper analyses the impact of credit expansions arising from decreases in collateral requirements or more expansionary monetary policies on long-term productivity in a model with endogenous growth. Credit expansions associated with relaxation of land collateral financing (capital collateral financing) will be productivity-and growth-retarding (enhancing). Without appropriate financial regulation, expansionary monetary policy may so encourage land speculation using leverage that productive capital investment is decreased; there is a temporary asset boom, but slower economic growth. The generation that experienced the asset price boom is better off, but subsequent generations are worse off because of low growth.

**Contact address:**

Tomohiro Hirano (Royal Holloway, University of London, Egham Hill, Egham TW20 0EX, UK), tomohih@gmail.com

Joseph E. Stiglitz (Columbia University, Columbia Business School 665 West 130th St.New York, NY 10027, US), jes322@columbia.edu

## Section 1. Introduction

### Section 1.1. Motivating empirical facts

Financial liberalization and expansionary monetary policy have been widely supported as leading to enhanced economic growth. The experience has been otherwise. Schularick and Taylor (2012), Jorda, Schularick, and Taylor (2013), Mian, Sufi, and Verner (2017), Greenwood et al. (2022), and Verner (2022) all demonstrate empirically that rapid expansion in credit systematically predicts growth slowdowns. The credit expansion of the early part of this century suggests an explanation: credit went disproportionately into real estate—so much so that the associated real estate boom crowded out more productive investments in other sectors.[1] Müller and Verner (2024) confirm that this is typical: using a novel database on the sectoral distribution of private credit for 117 countries starting in 1940, they show that credit expansions to the construction and real estate industries systematically predict subsequent productivity and economic slowdowns, while credit expansions to the manufacturing sector are associated with higher productivity and stronger macroeconomic performance. Moreover, Banerjee, Mehrotra, and Zampolli (2024) and Dai, Müller and Verner (2025) find empirically that with economic development, the share of real estate credit rises, and the price and reliance on real estate collateral increases, leading to a slowdown in long-term productivity and economic growth rates. On the other hand, productivity and economic growth tends to be higher in countries with a greater share of loans to the capital-intensive (manufacturing) sector.

### Section 1.2. The purpose of this paper

[1] Chakraborty, Goldstein, and MacKinlay (2018) empirically studied the effect of housing prices on bank commercial lending and firm investment in the U.S. in the period between 1988 and 2006. They found housing price booms led to crowding-out of commercial investment due to decreased lending by banks to credit-constrained firms. They conclude that housing price booms have negative spillovers to the real economy.

The purpose of this paper is to present a model that is consistent with these empirical findings. Key is endogenous growth where capital is the engine for economic growth, with spillovers to other factors of production (labor and land) or the other sectors (real estate).[2] We provide a parsimonious and tractable framework, an overlapping generations model with credit frictions, which enables us to contrast partial equilibrium effects with general equilibrium effects, and to capture the welfare effect of asset price booms depending on different generations. We will derive three main results

First, it is not so much aggregate credit expansion that matters for long-term productivity and economic growth but the type of credit expansions—where the credit goes—and that in turn depends on the varying degrees of collateral value across different assets. Credit expansions associated mainly with relaxation of real estate financing (of capital investment financing) will be productivity-and growth-reducing (enhancing) in the long run. This result, though consistent with the empirical findings noted in our motivating facts, is in sharp contract with the spirit of Kiyotaki and Moore (1997) and much of the modern macroeconomic/finance literature, which has suggested that increasing the value of land collateral (reducing collateral requirements) expands the opportunities for entrepreneurs to borrow to finance productive investments, increasing the efficiency of the macroeconomy. In our model, lowering collateral requirements might *seem* to enhance the ability of entrepreneurs to finance their investments *if* one were to ignore general equilibrium effects; but if it simultaneously results in an increase in land prices, the net effect is to "crowd out" productive investments, lowering growth, even though it generates temporary asset booms. We provide conditions under which the general equilibrium crowding out effect dominates.

[2] To illustrate our main results, we employ a one-sector endogenous growth model, but as the robustness of our analyses, in a later section 4, we show that our results hold even in a two-sector endogenous growth model.

Moreover, these results also imply that the macroeconomic impact of a real estate boom is markedly different between the short and long term. We show that although a real estate boom generates temporary consumption, output, and asset price booms, it reduces long-term productivity and economic growth, therefore lowering the wages and welfare of future generations.

Second, our model presents new insights into the consequences of monetary easing. The widely accepted view is based on Tobin (1965), where more expansionary monetary policy induces portfolio shifts from money to real capital, enhancing economic growth (see also Brunnermeier and Sannikov (2016) for a recent development). But in Tobin's world, there was no land, and so monetary easing leads to more capital investment. We show, by contrast, that without appropriate financial regulation, with lower interest rates and more expansionary monetary policy, more funds unevenly flow into the real estate sector rather than productive capital investment, which results in low growth and welfare, even though it generates temporary asset booms.[3] (This is a more general result: the macroeconomic consequences of policy can be markedly different between two-asset models of capital and money such as Tobin (1965) and our three-asset model of capital, money, and land.)

Third, the welfare impact of the low-interest-rate policy differs markedly depending on the generation that experienced the asset price boom and the subsequent generations with low growth. Those who hold land (asset-holder) at the time of the policy change receive unexpected capital gains, which increases their consumption and welfare. When the loan contract is nominal, an additional welfare-enhancing effect arises from a reduction in debt repayment due to unexpected increases in inflation rates, while the unexpectedly rising inflation decreases the welfare of the initial savers (lenders). Similarly, those who invest in land and capital using borrowing at that time also gain because the policy change leads to lower interest rate leading to high leverage. However, more expansionary monetary policy will result in low interest rates and low economic growth, thereby reducing the welfare of all future generations, regardless of their types.

[3] To describe monetary policy, we adopt Tobin's approach, which is based on the idea that economic agents hold various assets including money as part of a portfolio selection problem.

## Section 2. The model: endogenous growth with credit, money, and land

### 2.1 The environment

We use a variant of the standard two-period overlapping generations model where there is a competitive economy with productive capital, land and labor. Time is discrete and extends from zero to infinity ($t = 0, 1, \ldots$). In each period agents arrive and live for two periods. Each agent is endowed with one unit of labor when young, and supplies it inelastically, receiving wage income, $w_t$. A fraction $\eta$ of the young are entrepreneurs who have investment opportunities. The remaining fraction $1 - \eta$ are worker/savers (hereafter referred to as simply savers) who don't have investment opportunities. To simplify the analysis, all agents consume only in the second period, i.e., $u_t^i = c_{t+1}^i \geq 0$, where $u_t^i$ is the utility of agent $i$, and $i$ is the index for each agent.[4]

We focus on the case where only entrepreneurs participate in the real estate market. One interpretation of this assumption is that there is considerable land heterogeneity and savers are less informed about real estate, with high costs associated with obtaining the relevant information, making real estate unattractive for them. In section 4.2, we will examine the case where both agents participate in the real estate market and discuss further our assumption of limited asset market participation.

[4] In Appendix of our earlier working paper (Hirano and Stiglitz 2024), we also study a case in which each agent has log-utility over consumption in both young and old periods, respectively. We show that the results are unchanged. See Appendix of Hirano and Stiglitz (2024) for detail.

The budget constraint of entrepreneurial agent $i$ is given by

(1a) $k_{t+1}^i + P_t x_t^i + Q_t m_t^i = w_t + b_t^i$ and $c_{t+1}^i = R^c k_{t+1}^i + R_t^x P_t x_t^i + Q_{t+1} m_t^i - (1 + r_t) b_t^i$,

where $k_{t+1}^i$ and $x_t^i$ and $m_t^i$ are the entrepreneur's capital investment and land holdings and money holdings at date $t$. $b_t^i$ is the amount of borrowing at date $t$ if $b_t^i > 0$, and lending if $b_t^i < 0$. $R_t^c \equiv R_{t+1} + 1 - \delta$ is the total return per unit of capital investment made at date $t$, where $R_t$ is the rental rate of capital at date $t$, and $\delta$ is the depreciation rate of capital. $P_t$ and $D_{t+1}$ are the price of land at date $t$ and land rents at date $t + 1$ in terms of consumption goods. $Q_t$ is the price of money in terms of consumption goods ($1/Q_t$ is the price level). $R_t^x$ is the unleveraged return to land,

(1b) $R_t^x \equiv \frac{D_{t+1}}{P_t} + \frac{P_{t+1}}{P_t}$,

i.e., the return per dollar spent on land holdings without using borrowing between date $t$ and $t + 1$. It is a key endogenous variable, depending critically on $P_t$ and $P_{t+1}$.

Similarly, the budget constraint of savers can be written as

(1c) $Q_t m_t^i = w_t + b_t^i$ and $c_{t+1}^i = Q_{t+1} m_t^i - (1 + r_t) b_t^i + T_{t+1}^i$,

where $T_t^i$ is a lump sum transfer payment from the government to old savers.[5] Since savers don't have investment opportunities and don't participate in the real estate market, their behavior is simple, i.e., they lend their savings to entrepreneurs and/or hold money.

[5] We can also consider a case where government makes payments to old entrepreneurs. So long as $T_{t+1}^i$ cannot be pledged as collateral (e.g., social security payments are not allowed to be collateralized), our analysis applies as is; if it can be pledged as collateral, the analysis can be easily modified.

Since there is no risk associated with savers holding money or (by assumption) lending to entrepreneurs,[6] in equilibrium, the rates of returns must be the same.

(2) $1 + r_t = \frac{Q_{t+1}}{Q_t}$.

$\frac{Q_{t+1}}{Q_t}$ is the rate of return of a unit of money.[7]

We assume credit frictions. The entrepreneurs cannot borrow unless they have collateral. The borrowing constraint is

(3a) $(1 + r_t)b_t^i \leq \theta R_t^c k_{t+1}^i + \theta^x R_t^x P_t x_t^i$,

where the first and second terms of the right-hand side in (3a) reflect that only a fraction $\theta \in [0,1]$ of the total returns from capital investment and only a fraction $\theta^x \in [0,1]$ of the total returns from land holdings can be used as collateral. (3a) means that total debt repayment obligations cannot exceed the total value of the collateral (as evaluated at the time the loan is due).[8][9]

We also impose the short-sale constraint, i.e., agents can't hold a negative amount of money.

(3b) $m_t^i \geq 0$,

---

[6] If there is a finite support to the distribution of returns to entrepreneurial projects, the "zero" risk assumption can be translated into a maximum acceptable level of leverage.

[7] (2) implies that if $r_t > 0$, there is deflation. More generally, if there is a "convenience" return to holding money, $cr$, as is conventionally assumed, (2) becomes $1 + r_t = \frac{Q_{t+1}}{Q_t} + cr$, and so long as $r_t < cr$, the price level is increasing.

[8] See Stiglitz and Weiss (1986) and Hart and Moore (1994) for micro foundations of this type of debt contract. Of course, in this simplified model with no explicit risk, there are no defaults. In a more general model, even with collateral constraints, there are.

[9] In Appendix O3, we study the case with a different type of borrowing constraint (where the amount that can be borrowed relates to the entrepreneur's current assets) and show the robustness of our main results.

We now describe the production side of the economy. The productive sector exhibits standard Marshallian external increasing returns to scale in capital investment (Aoki 1970, 1971; Frankel 1962; Romer 1986.)[10] Capital is the engine for economic growth. Competitive firms produce the final consumption goods by using capital and labor. To keep things simple, we assume that the production function of each firm $j$ is given by

(4a) $y_{tj} = \left(k_{tj}\right)^{\alpha}\left(\chi(K_t)\, l_{tj}\right)^{\rho}\left(\chi(K_t)x_{tj}\right)^{\epsilon}$, with $\alpha + \rho + \epsilon = 1$.

Land, combined with capital and labor, is a necessary input for production.[11] $\chi(K_t)$ captures the positive spillovers on labor and land productivities, with $\chi'(K_t) > 0$, where $K_t$ is aggregate capital stock at date $t$. This formulation implies that productivities of labor and land grow at the same rate. This ensures the existence of a balanced growth path, as in standard growth theory. When we aggregate over all firms, we get at the aggregate level returns to scale. We assume $\chi(K_t)$ takes on a particular functional form:

(4b) $\chi(K_t) = aK_t$.

This is a key (though conventional) simplifying assumption.

[10] There is a large literature justifying the presence of these Marshallian increasing returns externalities, which we will not repeat; the existence of these effects is at the center of much of the endogenous growth literature. With scale economies, large scale enables high wages, which in turn has a feedback in allowing still larger scale, and with the right parameterization (the standard one, and the one we use here), this can generate steady endogenous growth. One big advantage of this approach is that it allows individual firms to face constant returns to scale—they ignore the aggregate scale benefit because their contribution to it is so small—while there are aggregate scale economies; thus, there can exist a competitive equilibrium, but of course, because of the externality, there is a presumption that it won't be efficient.

There is an alternative strand of research, growing out of Arrow's model of learning by doing (1962), where *improvements* in technology depend on gross investment today. That there are important differences should be clear: in the case of increasing returns, should something happen that would lead to less investment in the productive sector—say a decision by workers to consume more "housing" and less manufactured goods—then there could be a fall in productivity. By contrast, in the learning model, such a decision would simply slow the rate of increase of productivity. Productivity itself would not decline. Still, with appropriate parameterizations, one can get similar results out of the different formulations. (For examples of work in this tradition, see Stiglitz 1987, Stiglitz and Greenwald 2014, and Dasgupta and Stiglitz 1988).

[11] Nowadays, housing is also used as a factor of production. With this interpretation, land can be interpreted as housing.

## 2.2 The behaviour of individual firms

In the productive sector, the individual firm ignores its tiny influence on the aggregate capital stock and thus on the productivity of its own worker. Thus, each firm employs capital and labor up to the point where its private marginal product equals the rental rate of capital and the wage rate, respectively. Using (4b), factor prices and the aggregate production function become

(5a) $R_t = \alpha A$, (5b) $w_t = \rho A K_t$, (5c) $D_t = \epsilon A K_t$, and (5d) $Y_t = A K_t$,

where $A \equiv a^{\rho+\epsilon}$. This is a standard one-sector endogenous growth model with land. (5) holds at every momentary equilibrium, not just in steady state. The wage rate and aggregate output grow at the same rate of aggregate capital stock, so $\frac{w_{t+1}}{w_t} = \frac{K_{t+1}}{K_t} = \frac{Y_{t+1}}{Y_t}$ holds in steady state equilibrium, and the rental rate of capital is constant, with the increasing labor productivity as aggregate capital stock increases canceling out the effect of capital deepening and the diminishing returns associated with it.

## 2.3 The behavior of entrepreneurs and competitive equilibrium

We focus on the case where the borrowing constraint (3a) binds for entrepreneurs in equilibrium at each date, so that

(6) $R^c \equiv R + 1 - \delta = \alpha A + 1 - \delta > 1 + r_t$,

that is, the return on capital net of depreciation is greater than the safe rate of interest, implying that entrepreneurs would want to borrow as much as they could.[12] In this case, entrepreneurs would not hold money because the rate of return on money is strictly lower than the return on capital, i.e., the short-sale constraint (3b) will bind in equilibrium. As we will see below, (6) will hold in equilibrium if the money growth rate is positive.

[12] Indeed, each small entrepreneur, believing that (s)he has no effect on $R^C$ or $r$, would want to borrow an infinite amount to buy an infinite amount of the capital good. This, of course, ignores risk. Realistically, for many entrepreneurs, risk aversion is sufficiently low (or negative) that the binding constraint is the borrowing constraint that we focus upon. See our discussions in Appendix O1.

***Optimal portfolio allocations***

Entrepreneurs allocate their portfolio between capital and land, both using borrowing. Since they are perfect substitutes[13], in equilibrium, the leveraged rates of returns on each must be the same (see Appendix A1 for derivations):

$$\text{(7a)} \qquad \underbrace{\frac{R^c(1-\theta)}{1-\frac{\theta R^c}{1+r_t}}}_{\substack{\text{leveraged rate of return} \\ \text{on capital investment}}} = \underbrace{\frac{R_t^x(1-\theta^x)}{1-\frac{\theta^x R_t^x}{1+r_t}}}_{\substack{\text{leveraged rate of return} \\ \text{on land speculation}}} .$$

The no-arbitrage equation, (7a), is different from the standard one, and this difference is crucial for much of what follows. Solving (7a) for $R_t^x$ yields

$$\text{(7b)} \quad R_t^x = \frac{\lambda_t}{1-\theta^x+\frac{\theta^x\lambda_t}{1+r_t}},$$

where $\lambda_t \equiv \frac{R^c(1-\theta)}{1-\frac{\theta R^c}{1+r_t}}$ is the leveraged return on capital investment. For the denominator of the expression for the leveraged rate of return on capital investment to be positive, $1+r_t > \theta R^c = \theta(\alpha A + 1 - \delta)$.[14] Later, we will impose an assumption that ensures that that is the case. As we will show below, at the balanced growth path, (7b) can be solved for in terms of the basic technological and policy parameters of the model.

[13] The qualitative results would be similar if the two assets were imperfect substitutes, but the calculations would be much more complicated and less transparent.

[14] When (7a) holds, the equilibrium leverage associated with land speculation $1/\left(1-\frac{\theta^x R_t^x}{1+r_t}\right)$ is also positive and finite under Assumption 1, given below.

***The capital-investment function***

By substituting (2b) into (1a) and solving for $k_{t+1}^i$, we can derive the capital investment function of entrepreneurs when the borrowing constraint binds.

$$(8)\quad k_{t+1}^i = \underbrace{\frac{1}{1-\frac{\theta R^c}{1+r_t}}}_{\text{leverage}} \left[ \underbrace{w_t}_{\text{income}} - \underbrace{(1-\frac{\theta^x R_t^x}{1+r_t})P_t x_t^i}_{\substack{\text{total down-payment} \\ \text{in buying land}}} \right],$$

i.e. we can calculate an entrepreneur's capital holdings by taking his wage, subtracting what he has to pay to buy the land he holds (which depends on land leverage); and leveraging up that amount up through borrowing. $1-\frac{\theta^x R_t^x}{1+r_t}$ is the down-payment of a unit of land purchase.

***The competitive equilibrium***

The competitive equilibrium is defined as a set of prices $\{P_t, Q_t, w_t, R_t, R_t^x\}_{t=0}^{\infty}$ and quantities $\left\{c_t^i, b_t^i, x_t^i, m_t^i, k_{t+1}^i, \int c_t^i di, \int b_t^i di, \int x_t^i di, \int m_t^i di, \int k_t^i di, Y_t\right\}_{t=0}^{\infty}$, given an initial $K_0$, such that (i) each entrepreneur chooses land holdings, money holdings, capital investment, and borrowing to maximize utility under the budget and the borrowing constraints, and (ii) each saver lends his savings to entrepreneurs and/or hold money, and (iii) the market clearing conditions for land ($\int x_t^i di = 1$), money ($\int m_t^i di = M_t$), capital ($\int k_t^i di = K_t$), labor ($L_t = 1$), and goods ($\int c_t^i di + \int k_{t+1}^i di - (1-\delta)K_t = Y_t$) are all satisfied.

In the initial period $t = 0$, all land is held by the initial old entrepreneurs and all money is held by the initial old savers. We assume that the government expands the money supply at a fixed rate $\mu > 0$, so

$$(9)\quad M_{t+1} = (1+\mu)M_t,$$

where $M_t$ is money supply at date $t$.[15]

We assume the government does no borrowing, and thus must finance transfer payments out of money issuance, and the only transfer payments are made to old savers:

(10) $Q_t(M_t - M_{t-1}) = Q_t \mu M_{t-1} = \int T_t^i \, di.$

**2.4 Aggregate Dynamics**

We are now ready to derive aggregate dynamics. When we aggregate (8) across young entrepreneurs, we have

(11) $K_{t+1} = \frac{1}{1-\frac{\theta R^c}{1+r_t}}\left[\eta w_t - \left(1 - \frac{\theta^x R_t^x}{1+r_t}\right) P_t\right] = \frac{1}{1-\frac{\theta R^c}{1+r_t}}\left[\eta \rho A K_t - (1 - \frac{\theta^x R_t^x}{1+r_t}) P_t\right].$

To solve the dynamics, we introduce a variable $\phi_t \equiv \frac{P_t}{AK_t}$, the ratio of the value of land relative to aggregate production, i.e. GDP. For shorthand, we refer to this as the relative size of land speculation to GDP. From the definition of $\phi_t$, we can derive the evolution of $\phi_t$.

(12a) $\phi_{t+1} = \left(\frac{P_{t+1}/P_t}{1+g_t}\right) \phi_t,$

where $1 + g_t = \frac{K_{t+1}}{K_t}$, the growth rate of capital. The evolution of $\phi_t$ depends on the growth rate of output in the productive sector and that of land prices.

From the definition of $R_t^x$,

(13) $\frac{P_{t+1}}{P_t} = R_t^x - \frac{D_{t+1}}{P_t} = R_t^x - \frac{\epsilon A K_{t+1}}{P_t} = R_t^x - \frac{\epsilon A K_t}{P_t}\frac{K_{t+1}}{K_t} = R_t^x - \epsilon\left(\frac{1+g_t}{\phi_t}\right).$

[15] While in general, $\mu$ can vary period to period, in the steady state upon which we focus, $\mu$ is constant.

From (11), we can calculate the growth rate as a function of $\phi_t$ and $r_t$ and the various parameters of the problem:

(14a) $$1 + g_t = \frac{A}{1-\frac{\theta R^c}{1+r_t}}\left[\eta\rho - (1 - \frac{\theta^x R_t^x}{1+r_t})\phi_t\right].$$

Other things being constant, a rise in $\phi_t$ (land speculation) crowds out savings away from capital investment, reducing the growth rate of the economy. On the other hand, an increase in $\theta$ or $\theta^x$, or a fall in $r_t$ increases the leverage of capital investment and decreases the down-payment of a unit of land purchase, generating more capital investment. We say that capital investment is "crowded in". The question is what are the circumstances under which each effect dominates? Substituting (13) and (14a) into (12a) yields

(12b) $$\phi_{t+1} = \frac{\phi_t R_t^x}{1+g_t} - \epsilon = \phi_t R_t^x / \left\{\frac{A}{1-\frac{\theta R^c}{1+r_t}}\left[\eta\rho - (1 - \frac{\theta^x R_t^x}{1+r_t})\phi_t\right]\right\} - \epsilon.$$

The value of money is determined so that savings each period (the income of young people, both entrepreneurs and savers) equals holdings of capital, land, and money:

(15a) $$Q_t M_t = w_t - K_{t+1} - P_t.$$

Using the definition of $\phi_t$, (15a) can be written as

(15b) $$Q_t M_t = \left\{\rho - \frac{1}{1-\frac{\theta R^c}{1+r_t}}\left[\eta\rho - (1 - \frac{\theta^x R_t^x}{1+r_t})\phi_t\right] - \phi_t\right\} A K_t,$$

So long as the term in the large bracket is positive, the price of fiat money will be positive. Intuitively, if the savings remaining after financing capital investment and land holdings are large enough, those idle savings flow toward money holdings.

The dynamics of this economy, entailing $\{P_t, Q_t, K_t, M_t, r_t, T_t\}$ can be characterized by (2), (9), (10), (12b), (14a) and (15b). In spite of the seeming complexity, it turns out that the dynamics are quite simple: There is a unique steady state, and the economy immediately converges to it.

### 2.5 Steady State

In steady state the real value of money grows according to

(16) $1 + r^* = \frac{Q_{t+1}}{Q_t} = \frac{1+g^*}{1+\mu}$.

As is standard in this class of models, whether inflation or deflation occurs in equilibrium depends on the relative size of $g^*$ and $\mu$. (12b), (14a), and (16) give three equations in three unknowns, $r^*$, $g^*$, and $\phi^*$.

***Steady states: unproductive land*, i.e., $\epsilon = 0$**

We first study a special case of $\epsilon = 0$, i.e., unproductive land. In this case, $\rho = 1 - \alpha$. Even if land is unproductive, it can still be used as collateral and as a store of value. The steady-state growth rate is $1 + g^* = R^{x*} = \frac{\lambda}{1-\theta^x + \frac{\theta^x \lambda}{1+r^*}}$, where it will be recalled $\lambda \equiv \frac{R^C(1-\theta)}{1-\frac{\theta R^C}{1+r^*}}$. Using (16), we have $(1 + r^*)(1 + \mu) = \frac{\lambda}{1-\theta^x+\frac{\theta^x \lambda}{1+r^*}}$, which, after much simplifying, can be solved for $r^*$, and hence we can now solve for $g^*$ and $\phi^*$ explicitly:

(17a) $1 + r^* = \frac{R^C}{1-\theta^x}\left[\frac{1-\theta}{1+\mu} - (\theta^x - \theta)\right]$.

(17b) $\phi^* = \frac{\eta(1-\alpha)}{1-\theta^x(1+\mu)} - \frac{R^C(1-\theta)}{A(1-\theta^x)}$.

(17c) $1 + g^* = \frac{R^C}{1-\theta^x}[1 - \theta - (\theta^x - \theta)(1 + \mu)]$.

i.e., steady state growth depends in a simple way on the leverage variables, the rate of growth of money, and the total return per unit of capital investment. Assumption 2 below guarantees that $\phi^*$ is positive. Note that there is another steady state with $\phi_t = 0$ for all $t$—effectively a "landless" economy. But as we will see below, such a "landless" equilibrium cannot exist with positive $\epsilon$.

***General case***

The more general case is not solved so simply. By rearranging (12b), (14a), and (16), we can derive the quadratic equation for $\phi^*$ (see Appendix A2 for detail). We can show there that when $\epsilon > 0$, one of the solutions is positive under Assumption 2 we introduce below, and the other one is negative (when $\epsilon = 0$, the other one is zero). Solving $\phi^* > 0$ yields

$$(18b) \quad \phi^* = \frac{-\Upsilon_2 + \sqrt{\Upsilon_2^2 + 4A[1-\theta^x(1+\mu)]\epsilon\left\{\eta\rho A + A\theta^x(1+\mu)\epsilon + \frac{R^C(1-\theta)}{1-\theta^x}\theta^x(1+\mu)\right\}}}{2A[1-\theta^x(1+\mu)]},$$

where

$$\Upsilon_2 = -\eta\rho A - A\theta^x(1+\mu)\epsilon + A\epsilon[1-\theta^x(1+\mu)] + \frac{R^C(1-\theta)}{1-\theta^x}[1-\theta^x(1+\mu)].$$

(17a) and (17c) can also be written as

$$(18a) \quad 1 + r^* = \frac{R^C}{1-\theta^x}\left[\frac{1-\theta}{(1+\mu)(1+\frac{\epsilon}{\phi^*})} - (\theta^x - \theta)\right],$$

$$(18c) \quad 1 + g^* = (1+r^*)(1+\mu) = \frac{R^C}{1-\theta^x}\left[\frac{1-\theta}{(1+\frac{\epsilon}{\phi^*})} - (\theta^x - \theta)(1+\mu)\right].$$

The critical variables, $g^*$, $r^*$, and $\phi^*$, are functions of the parameters $\{a, \alpha, \epsilon, \eta\}$ and the policy variables $\{\theta^x, \theta, \mu\}$.

## 2.6 Existence of equilibrium

In the rest of our analysis, to obtain clear economic intuition, we focus on the case with $\epsilon$ sufficiently small, i.e. it captures the decline in the role of land as a factor of production.[16] Note that if $\epsilon \to 0$, $\rho \to 1-\alpha$.

To ensure the existence of equilibrium, we make the following assumptions. For the equilibrium leverage associated with capital investment to be finite, we impose a constraint on the maximum rate of money growth.

**Assumption 1.** $1 > \theta^x(1+\mu)$, or $1+\mu < 1/\theta^x$[17].

We also impose constraints on parameter values to ensure $\phi^* > 0$ and $Q_t > 0$. A necessary and sufficient condition for these inequalities to hold simultaneously is

**Assumption 2.** $A(1-\alpha) + \frac{R^c(\theta^x-\theta)(1+\mu)}{1-\theta^x} > \frac{\eta(1-\alpha)A}{1-\theta^x(1+\mu)} > \frac{R^c(1-\theta)}{1-\theta^x}$.

The inequality on the left-hand side ensures $Q_t > 0$, i.e., money circulates in equilibrium,[18] and the inequality on the right-hand side ensures $\phi^* > 0$, where $\phi^*$ is given by (18b). We can easily find that there are relevant parameter values for which Assumptions 1 and 2 simultaneously hold.[19]

---

[16] Indeed, as a historical trend, as economies develop, this has been the case. Our earlier working paper (Hirano and Stiglitz 2024) presents numerical analyses for the case of large $\epsilon$.

[17] If the rate of increase of the money supply is too high, the interest rate gets sufficiently low that equilibrium leverage will become infinite.

[18] It is known that in a standard model with land, $1+r > 1+g$ holds at a balanced growth path, otherwise, land prices would be infinite. If $1+r > 1+g$ holds, money cannot circulate. However, in our model, since entrepreneurs invest in capital and land, both using borrowing, and the no-arbitrage equation (7a) is different from a standard one, $1+r_t = \frac{D_{t+1}}{P_t} + \frac{P_{t+1}}{P_t}$. Therefore, money can circulate in equilibrium.

[19] Note that the inequality $A(1-\alpha) + \frac{R^c(\theta^x-\theta)(1+\mu)}{1-\theta^x} > \frac{R^c(1-\theta)}{1-\theta^x}$ can be written as

We can also verify that the borrowing and short-sale constraints, (2a) and (2b), bind in equilibrium, i.e., using (18a), (6) can be written as $(1+\mu)\left(1+\frac{\epsilon}{\phi^*}\right)>0$, which is true under Assumptions 1 and 2 if $1+\mu>0$.

We obtain the following Proposition.

**Proposition 1 Uniqueness of balanced growth path**

*Under Assumption 1 and 2, there exists a unique balanced growth path where the relative size of land speculation $\phi^*>0$, economic growth rate $1+g^*$, the interest rate $1+r^*$, which equals the rate of return on money $\frac{Q_{t+1}}{Q_t}$, are constant over time.*

### 2.7 Comparative steady state analysis

(17a), (17b), and (17c) allow us to conduct comparative statics with changes in the market/policy parameters. We are particularly interested in impacts on growth. We can see how growth is affected by changes in the interest rate or collateral requirements by decomposing the effects of changes into a direct (partial equilibrium) effect and a general equilibrium effect, where the latter represents the impact on the equilibrium level of speculation, through (12b):

$$(14b)\quad 1+g^* = \underbrace{\frac{1}{1-\frac{\theta R^C}{1+r^*}}}_{\text{PE effect 1}}\left(\eta(1-\alpha)-\underbrace{\left(1-\frac{\theta^x R^{x*}}{1+r^*}\right)}_{\text{PE effect 2}}\underbrace{\phi^*}_{\text{GE effect}}\right)A.$$

$\left(1-2\alpha-\frac{1-\delta}{A}\right)(1-\theta^x)+\left(\alpha+\frac{1-\delta}{A}\right)\mu(\theta^x-\theta)>0$. An easily satisfied set of sufficient conditions for this to hold is that $\left\{A>\frac{1-\delta}{1-2\alpha},\alpha<\frac{1}{2}\text{ and }\theta^x\geq\theta\right\}$. By choosing $\eta$ appropriately and ensuring that $1+\mu<1/\theta^x$, we can verify that there are relevant parameters for which Assumptions 1 and 2 are simultaneously satisfied.

where it will be recalled $R^{x*} = \frac{\lambda}{1-\theta^x+\frac{\theta^x\lambda}{1+r^*}}$, where $\lambda \equiv \frac{R^C(1-\theta)}{1-\frac{\theta R^C}{1+r^*}}$. PE effect 1 represents the direct leverage effect, with investment and growth increasing with a rise in $\theta$ or with a decline in $r^*$. PE effect 2 represents the effect on capital investments through changes in the equilibrium down-payment in buying a unit of land. With increases in $\theta$[20] or $\theta^x$ or with a decrease in the interest rate caused by an increase in $\mu$, the equilibrium down-payment in buying land decreases, allowing the financing of more capital investments. On the other hand, the GE effect represents the effect on capital investments as a result of changes in the relative size of land speculation. The various effects sometimes pull in different directions; nonetheless, we are able to establish:

**Proposition 2 Impact of changes in the collateral values on long-run economic growth in a monetary economy**

*(2-i) For $\epsilon$ sufficiently small, $\frac{d(1+g^*)}{d\theta^x} < 0$, with $\frac{d\phi^*}{d\theta^x} > 0$[21] and $\frac{d(1+r^*)}{d\theta^x} < 0$.*

*That is, with greater $\theta^x$, land speculation increases, which produces a crowding-out effect. On the other hand, the interest rate declines,[22] which increases equilibrium leverage with capital investment and decreases the down-payment on a unit of land, producing a crowding-in effect. Nonetheless, the crowding-out effect dominates the crowding-in effect: the increase*

[20] If $\theta$ increases, a portfolio shift will occur from land speculation to capital investment. For the leveraged return on land and capital to remain equal, $R^{x*}$ must increase. As the unleveraged rate of return on a unit of land increases, it raises the collateral value of land, thereby reducing the equilibrium down-payment in buying a unit of land.

[21] From (17b), we have $\frac{d\phi^*}{d\theta^x} = \frac{\eta(1-\alpha)(1+\mu)}{[1-\theta^x(1+\mu)]^2} - \frac{R^C(1-\theta)}{A(1-\theta^x)^2} > 0$ if and only if $\frac{\eta(1-\alpha)(1+\mu)}{[1-\theta^x(1+\mu)]^2} > \frac{R^C(1-\theta)}{A(1-\theta^x)^2}$. This inequality condition can be written as $\frac{\eta(1-\alpha)}{[1-\theta^x(1+\mu)]} > \frac{(\alpha A+1-\delta)(1-\theta)}{A(1-\theta^x)}\frac{[1-\theta^x(1+\mu)]}{(1-\theta^x)(1+\mu)}$, which holds under Assumption 2 because $\frac{[1-\theta^x(1+\mu)]}{(1-\theta^x)(1+\mu)} < 1$ if $\mu > 0$.

[22] From (16), with a decline in $g^*$, interest rates have to decline if there is to be a steady state.

*in land speculation is so great that productivity and economic growth are impaired even though the interest rate has decreased.*

*(2-ii) For* $\epsilon$ *sufficiently small,* $\frac{d(1+g^*)}{d\theta} > 0$*, with* $\frac{d\phi^*}{d\theta} > 0$ *and* $\frac{d(1+r^*)}{d\theta} > 0$*.*

*That is, an increase in the collateral value of capital investment increases leverage, which finances more capital investment, thereby increasing productivity and economic growth, even though there is an endogenous rise in the interest rate and in our measure of relative land speculation* $\phi^*$*.*[23]

Next, we examine the effect of low-interest-rate policy. As $\epsilon \to 0$, we obtain

$$(19)\quad 1 + g^* = R^{x*} = \frac{\lambda}{1-\theta^x+\frac{\theta^x\lambda}{1+r^*}} = \frac{R^c(1-\theta)(1+r^*)}{(1-\theta^x)(1+r^*)+R^c(\theta^x-\theta)}.$$

Then, differentiating (19) with regard to $r^*$ directly leads to the following Lemma.

**Lemma 1** *For* $\epsilon$ *sufficiently small,* $sign\frac{d(1+g^*)}{dr^*} = sign(\theta^x - \theta)$*.*

That is, the long-run impact of low interest rates on economic growth depends on the relative size of $\theta^x$ and $\theta$. When $\theta^x > \theta$, the crowding-out effect caused by low interest rates (with land prices increasing) dominates the crowding-in effects; low interest rates encourage land speculation with leverage more than capital investment, reducing long-run productivity and economic growth. On the other hand, when $\theta^x < \theta$, just the opposite is true.

[23] This occurs because the increased leverage in capital investment allows more funds to be available for land speculation.

Moreover, from (17a) and (17b), we know $\frac{d(1+r^*)}{d\mu} < 0$ and $\frac{d\phi^*}{d\mu} > 0$. That is, an increase in money growth rate reduces the interest rate and increases the relative size of land speculation. Then, these results and Lemma 1 lead to the following Proposition.

**Proposition 3 Impact of monetary policy on long-run economic growth**

*For $\epsilon$ sufficiently small, $sign\frac{\partial(1+g^*)}{\partial\mu} = sign(\theta - \theta^x)$. That is, the long-run impact of expansionary monetary policy (keeping all other parameters constant) on economic growth depends on the relative size of $\theta^x$ and $\theta$.*

A loosening of monetary policy or of collateral requirements may so increase land speculation that resources are diverted from capital investment and growth may be impaired—consistent with the experiences in numerous countries and the studies cited in the introduction. To ensure that easier credit leads to more investment, at the same time as monetary easing, financial regulations on the real estate sector need also be introduced, i.e., $\theta > \theta^x$ so that more funds can flow into capital investment rather than land speculation.

From Propositions 2 and 3, we can derive the following Corollary.

**Corollary** *The land price-rent ratio ($\frac{P_t}{D_t}$), the aggregate value of land-GDP ratio ($\frac{P_t}{Y_t}$), the wealth-GDP ratio ($\frac{Y_t+P_t}{Y_t}$), and the Credit/GDP ratio ($\frac{B_t}{Y_t}$) all rise as $\theta$ or $\theta^x$ or $1+\mu$ increases,[24] where we define GDP at date $t$ as $Y_t$ (defined in (4d)).*

---

[24] Let us explain how the Credit/GDP ratio responds to policy changes in more detail. When the borrowing constraint binds, the Credit/GDP ratio on the balanced growth path can be written as $\frac{B_t}{Y_t} = \frac{\theta R^c}{1+r^*}\frac{K_{t+1}}{Y_t} + \frac{\theta^x R^x}{1+r^*}\frac{P_t}{Y_t}$. Since $1+r^* = \frac{1+g^*}{1+\mu}$ and $\frac{\theta^x R^x}{1+r^*} = \theta^x(1+\mu)$ on the balanced growth path, the ratio can be written as $\frac{B_t}{Y_t} = \frac{\theta R^c(1+\mu)}{A} + \theta^x\phi^*(1+\mu)$. Since we know from (17b) that $\phi^*$ is increasing functions of $\theta$ or $\theta^x$ or $1+\mu$ for small $\epsilon$, the Credit/GDP ratio rises as $\theta$ or $\theta^x$ or $1+\mu$ increases.

Two remarks are in order. First, in equilibrium, since GDP at date $t$ equals aggregate consumption and aggregate capital investments at date $t$, Proposition 2-i and 3 imply that with a low-interest-rate policy or an increase in land leverage, the aggregate consumption/GDP ratio increases, while the aggregate net investments/GDP ratio declines. Moreover, with more expansionary monetary policy or an increase in land leverage, land prices rise instantly, generating temporary asset price booms accompanied by increased aggregate consumption (see also footnote 30).

Second, although we have focused on the case of $\epsilon \to 0$, if we employ a different type of borrowing constraints, which is $b_t^i \leq \theta k_{t+1}^i + \theta^x P_t x_t^i$, i.e., this collateral constraint is set as a fraction of this year's value of land, then, regardless of the size of $\epsilon$, we can analytically show $\frac{d(1+g^*)}{d\theta^x} < 0$, $\frac{d(1+g^*)}{d\theta} > 0$, and $sign\frac{d(1+g^*)}{d\mu} = sign(\theta - \theta^x)$. (See Appendix O2.) Intuitively, because the interest rate is not involved with the constraint, changes in $r_t$ do not impact the crowding-in PE effects 1 and 2. Therefore, when $\theta^x > \theta$, low interest rates caused by expansionary monetary policy end up increasing the size of land speculation, thereby negatively impacting long-run economic growth.

***Comparison with the standard view of more expansionary monetary policy (Tobin (1965))***

Proposition 3 is in sharp contrast with the conventional view of expansionary monetary policy in which there is a presumption that easier monetary and credit policies (e.g. a faster rate of expansion of the money supply or lower collateral requirements) lead to more capital investment. See, e.g. Tobin (1965) and Brunnermeier and Sannikov (2016). Tobin claimed that an increase in the money growth rate leads to higher inflation rates and reduced rates of return on money, inducing a portfolio shift from money to real capital, thereby leading to higher long run economic growth. This effect is called the "Tobin Effect".[25]

Tobin (and much of the related subsequent literature on money and growth) was correct that higher rates of money growth lead to higher rates of inflation; but he ignored the existence of land, and the shift away from money may (and often does) lead to a demand for more land rather than more real investment. Our model can be used to validate Tobin's claim in a landless economy, i.e. one where $\epsilon = 0$ and $P_t = 0$. From (14a), when $P_t = 0$, at the balanced growth path, the growth rate of the economy can be written as

$$(20)\quad 1 + g^* \equiv \frac{K_{t+1}}{K_t} = \frac{\eta(1-\alpha)A}{1 - \frac{\theta R^c}{1+r^*}}.$$

It is clear from (20) that a reduction in the interest rate relaxes the borrowing constraint, financing more capital investment and leading to higher economic growth. By substituting (20) into (16) and solving for $1 + r^*$, we have the standard result.

$$(21)\quad 1 + r^* = \frac{Q_{t+1}}{Q_t} = \frac{\eta(1-\alpha)A}{1+\mu} + \theta R^c.$$

That is, an increase in the money growth rate increases the inflation rate and lowers the long run interest rate. We summarize this result as the following Proposition.

**Proposition 4 Replication of the Tobin's claim in a model with credit frictions**

*Consider a special case with $\epsilon = 0$ and $P_t = 0$, i.e., the landless economy. Then, for $\theta > 0$, we have $\frac{d(1+g^*)}{d\mu} > 0$, with $\frac{d(1+r^*)}{d\mu} < 0$.*

[25] There is, of course, a large literature identifying other effects on higher inflation, arguing that it is bad for economic growth. The simple growth model we present here is consonant with that investigated by Tobin, as we show below.

Because decreased interest rates relax the borrowing constraint of entrepreneurs, more capital investment is financed and productivity and economic growth are higher.**2.8 Dynamics**

Looking at the difference equations for $\{P_t, K_t, Q_t, M_t, r_t, T_t\}$ [defined by (2), (9), (10), (12b), (14a) and (15b)], it is clear that if the price of land at time 0 jumps to $P_0 = AK_0\phi^*$, and the price of money jumps to $Q_0 = \left\{\rho - \frac{1}{1-\frac{\theta R^c}{1+r*}}\left[\eta\rho - \left(1 - \frac{\theta^x R^x}{1+r^*}\right)\phi^*\right] - \phi^*\right\}\frac{AK_0}{M_0}$, then there is a steady state equilibrium where all the dynamical equations are satisfied for all $t \geq 0$.

More generally, we show in Appendix A3 that the dynamics can be reduced to the two-dimensional dynamical system, with $\phi_{t+1} = \psi(\phi_t, 1 + r_t)$ and $1 + r_{t+1} = \omega(\phi_t, 1 + r_t)$, where both $\phi_t$ and $1 + r_t$ are variables that can jump in value. We can show that there is a dynamic equilibrium path in which $\phi_t$ and $1 + r_t$ jump to their steady state value at $t = 0$. This rational expectations path always exists under our Assumptions 1 and 2.

**Proposition 5**: **Existence of rational expectations path without transitional dynamics**

*There exists a rational expectations path along which the economy achieves the balanced growth path immediately without transitional dynamics.*[26]

### 2.9 Welfare implications

Changes in the market/policy parameters $\{\mu, \theta^x, \theta\}$ can affect agents of different generations and types differently. Consider a low-interest-rate policy that begins at time $T$, i.e. the government announces that going forward, they will increase the money growth rate $\mu$ faster than had previously been the case. Suppose the case of $\theta^x > \theta$. This policy change results in the price of land going up initially at time $T$ but the growth rate slowing from time $T$ onwards. Moreover, the real value of money decreases initially, and the inflation rate increases. We can show that nonetheless, the value of the assets of the elderly increases, and the productive investments of the young decrease, so that consumption of the elderly increase—they are better off.

[26] In the different borrowing constraint we study in Appendix O2, we can prove that there exists a unique rational expectations path along which the economy achieves the balanced growth path immediately without transitional dynamics and this is the only possible rational expectations trajectory.

To derive this, we use (10) to establish that

$$(22)\quad Q_T = \left\{\rho - \frac{1}{1-\frac{\theta R^c}{1+r_T}}\left[\eta\rho - \left(1 - \frac{\theta^x R^x}{1+r_T}\right)\phi_T\right] - \phi_T\right\}\frac{AK_T}{M_T},$$

Given $K_T$ and $M_T$, with more expansionary monetary policy, $Q_T$ (the price of money in terms of consumption goods) decreases, i.e. the inflation rates rises unexpectedly, but the total wealth of the elderly $Q_T M_T + P_T = Q_T M_T + AK_T\phi_T$ has increased.[27] The elderly entrepreneurs are better off because they receive capital gains which increases their consumption; and with national output fixed, that means productive investment decreases—a result which also follows from the fact that with fixed savings, more of savings goes into holding non-produced assets.

The entrepreneurs of the first generation ($q = 0$) are better off because the leveraged rate of return increases due to the decline in $1 + r_T$. The welfare of entrepreneurs from the date $q = 1$ onwards depends on two opposing factors: the positive effect of increased leverage and the negative effect of the slowdown in wages. Although entrepreneurs in the first few generations may be better off or worse off, depending on the relative strength of these opposing effects, entrepreneurs far enough into the future (for $q$ large enough) will be surely worse off because the wage decline will dominate.

[27] Using (22), we obtain $Q_T M_T + AK_T\phi_T = \rho - \frac{1}{1-\frac{\theta R^c}{1+r_T}}\left[\eta\rho - \left(1 - \frac{\theta^x R^x}{1+r_T}\right)\phi_T\right] = \rho - \frac{1+g^*}{A}$. Since $1 + g^*$ decreases with an increase in $\mu$ if $\theta^x > \theta$, the total wealth of the elderly increases, while it decreases if $\theta^x < \theta$.

There are three effects on the utility of savers born after time $T + q$ ($q \geq 0$): It is lowered because $g^*$ is lowered, so wages at each date are lowered, and $r_{T+q}$ will decrease. But the increase in $\mu$ changes the real value of the amount of lump sum transfer, $(1 + \mu)Q_{T+q}M_{T+q}$, and changes the rate at which it grows. In particular, we note that for balanced growth, the rate of increase in transfers (which must equal $g^*$) has been lowered, and that, together with the lower rate of increase in wages combined with the lower interest rate, implies that eventually savers are worse off.

**Proposition 6 The welfare implications of a low-interest-rate policy**

*Without financial regulation, a low-interest-rate policy (more expansionary monetary policy)* ***(i)*** *benefits the initial elderly entrepreneurs, who hold land, and the initial young entrepreneurs, who invest in land and capital using borrowing, but* ***(ii)*** *it will decrease the welfare of the initial elderly savers and the welfare of all savers*[28] *and entrepreneurs far enough in the future.* ***(iii)*** *Entrepreneurs in the first few generations after the policy change may be better off or worse off, depending on the welfare-increasing effect from the increased leveraged return and the welfare-decreasing effect resulting from the slowdown in wages. So too for savers in the first few generations, i.e., the declines in the interest rate and wages and the rise in inflation rates decrease their welfare, while the increase in the direct lump sum transfer increases their welfare.*

[28] Due to the policy change, the inflation rate rises unexpectedly, which decreases the rate of return on money holding and the return from lending, which lowers savers' consumption and welfare.

We can obtain a similar welfare impact when land leverage $\theta^x$ increases, as this leads to reduced growth and lower interest rates.

There is a message to be drawn from this analysis. Although more expansionary monetary policy generates temporary asset price booms, benefiting the initial asset holders and borrowers, the policy, without financial regulation, will result in low interest rates, slow economic growth, and reduced welfare for future generations, implying markedly different welfare implication between the generation who experienced the asset price boom and the subsequent generations with low growth.

**Section 3**. **Equilibrium with finite land prices, low interest rates, and high returns to capital**

In standard macroeconomic theory, if $r$ is very low, in particular less than the growth rate, there is a problem. There is dynamic inefficiency. Such an equilibrium can easily arise in life-cycle models without land. But if there exists an asset like land (a non-reproducible asset), with balanced growth no matter how small, as $r$ goes to zero, its market value becomes infinite, which of course cannot be an equilibrium. The only possible equilibria entail $r > g$.

But it has been the norm in the U.S economy that the return on government bonds has been less than the growth rate. The model we propose is at least theoretically consistent with this empirical observation.

Along the balanced growth path, we know that $1 + r^* = \frac{1+g^*}{1+\mu}$. It is obvious that $1 + g^* > 1 + r^*$ if $\mu > 0$. We also know from (14c) that $1 + g^* = R^{x*}/[1 + (\epsilon/\phi^*)]$. So long as $\epsilon > 0$, $R^{x*} > 1 + g^*$. Moreover, $R^c > R^{x*}$ if $\theta^x > \theta$. Therefore, if $\theta^x > \theta$ and $\mu > 0$, $R^c > 1 + g^* > 1 + r^*$ holds along the balanced growth path. When $\epsilon \to 0$, $1 + g^* \to R^{x*}$ and the same argument applies. We summarize this result as the following Proposition.

**Proposition 8 Finite land prices with low interest rates**

*If* $\theta^x > \theta$, $\epsilon \geq 0$, *and* $\mu > 0$, $the\ return\ to\ capital > 1 + g^* > 1 + r^*$ *holds along the balanced growth path and* $P^*$ *is finite.*

## Section 4. Discussions

### Discussion 1. Concerning the assumption of limited asset market participation

There is a good justification for our assumption of limited asset market participation by one part of the population, those we have referred to as workers. There is considerable land heterogeneity and workers are less informed about it (with high costs associated with obtaining the relevant information), making land unattractive for them. For instance, there are real estate ventures yielding a zero return and those yielding a large positive return, and because non-entrepreneurial workers can't distinguish between them, they do not invest in land, especially if they are risk averse; the expected return is just too low.

If there are a finite number of projects, the informed will select the good projects (on average), leaving a disproportionate share of bad projects for the uninformed, so that even if the expected returns of the informed is equal to or greater than other investment opportunities, that of the uninformed can be less even than government bonds.

There are other reasons for limited participation in the land market. For instance, people have different attitudes toward risk. The land market is obviously more risky than the (real) return on government bonds, so very risk averse agents don't participate in the land market. In practice, workers, having lower incomes than entrepreneurs, may be more risk averse.

It should be obvious that this assumption is crucial. If both entrepreneurs and savers participate in the land market, then, the no-arbitrage condition will be changed to $1 + r_t =$ $\frac{D_{t+1}}{P_t} + \frac{P_{t+1}}{P_t}$ (a standard one), in which case the discount rate will be $1 + r^*$ on the steady-state growth path. If this is the case, at the steady-state growth path, the land price-rent ratio can be written as

$$\frac{P_t}{D_t} = \frac{1+g^*}{1+r^*-(1+g^*)}.$$

In the case where $1 + g^* > 1 + r^*$, which holds empirically over the past century of U.S data (see Blanchard (2019)), the ratio would become infinite and hence, there is no equilibrium in the model as we have formulated it.

Moreover, if we substitute (16) giving the interest rate as a function of the growth rate of money and capital into the above price-dividend ratio, we obtain

$$\frac{P_t}{D_t} = \frac{1+g^*}{\frac{1+g^*}{1+\mu}-(1+g^*)}.$$

For the existence of equilibrium, $\mu < 0$, i.e., the growth rate of money would have to be negative, otherwise no equilibrium exists. Of course, in the real world, $\mu > 0$, i.e., the money supply has expanded.

**Discussion 2. Description using a two-sector growth model**

To illustrate the key idea, we have employed a one-sector endogenous growth model. Our analysis can be extended to include two sectors, a productive sector, which in our simplified two-sector setting one can think of as manufacturing (in the model of the text, the only sector that uses capital; more generally, the capital-intensive sector) and a real estate sector (which in the simplified version of our model, only uses land). The production function of each firm $j$ is now given by

(23a) $y_{tj} = (k_{tj})^{\alpha}(\chi(K_t)l_{tj})^{1-\alpha}$,

where $k_{tj}$ and $l_{tj}$ are capital and labor inputs of firm $j$ in the capital-intensive sector. $\chi(K_t)$ is given by (4b).

In the real estate sector, one unit of land produces $D_t$ units of consumption goods (in Appendix O3, we relax this assumption and examine a case where both labor and land are inputs for real estate). To ensure the existence of the balanced growth path, we focus on a case where there is a full spillover from the productive sector to the real estate sector, so that land productivity grows at the economy's growth rate,[29]

(23b) $D_t = \nu\chi(K_t) = \nu a K_t$,

where $\nu \geq 0$ is a parameter that captures the extent to which an increase in labor productivity in the capital-intensive sector spills over to an increase in land productivity, and refer to it as the spillover coefficient; when $\nu$ is larger, with an increase in $K_t$, the productivity increase in the real estate sector is larger.[30] Aggregate output $Y_t$ is then $Y_t = (K_t)^{\alpha}(\chi(K_t)L_t)^{1-\alpha} + D_t X_t$.[31]

[29] Appendix O3 in our working paper version (Hirano and Stiglitz 2024) also studies the case with imperfect spillovers and show the robustness of our main results.
[30] So long as the positive spillover effect on the real estate sector is a linear function of $K_t$, we can ensure the existence of a balanced growth path. Our approach to achieving balanced growth is similar to that taken by Greenwald and Stiglitz (2006). More generally, to obtain balanced growth, we need to construct a model so that different sectors will at least eventually grow at the same rate. It is clear from our parametrizations that the rate of growth of productivity in the productive sector equals the rate of growth of productivity in real estate.
[31] As another interpretation, this production function corresponds to the limiting case of the CES production function, $Y_t = \left(\gamma_1((K_t)^{\alpha}(\chi(K_t)L_t)^{1-\alpha})^{\frac{\sigma-1}{\sigma}} + \gamma_2(\chi(K_t)X_t)^{\frac{\sigma-1}{\sigma}}\right)^{\frac{\sigma}{\sigma-1}}$, where $\sigma$ is the elasticity of substitution between the manufacturing and real estate sectors [between $(K_t)^{\alpha}(\chi(K_t)L_t)^{1-\alpha}$ and $\chi(K_t)X_t$], and $\gamma_1$ and $\gamma_2$ are parameters and are set to be unity, as $\sigma \to \infty$.

Using this two-sector description, $\rho$ and $\epsilon$ in sections 2 and 3 are now replaced with $1-\alpha$ and $\nu a^{\alpha}$, respectively, and $\nu \to 0$ corresponds to $\epsilon \to 0$. All analyses are the same as in sections 2 and 3.

With this two-sector growth model, sectoral credit expansion — whether it is based on credit expansions to the capital-intensive sector or to the real estate sector— matters to long-term productivity and economic growth.

**Temporary output and asset price booms and their long-lasting negative effects on future generations**

Using this two-sector description, we can analyze the short-run and long-run effects on the economy of a real estate boom brought on by an increase in productivity in the real estate sector. Normally, one would have thought that an increase in productivity—an expansion of today's production possibilities curve—is unambiguously welfare enhancing. It turns out that that is not the case.

Suppose that at the beginning of date $s$, there is an unexpected *permanent* increase in $\nu$, generating a real estate boom. With this shock, the size of land speculation increase, while the growth rate of aggregate capital decreases (see Appendix A4 for the derivations). This means that land prices at date $s$, $P_s$, rise, given $K_s$. Since aggregate output and consumption at date $s$, $Y_s$ and $\int c_t^i di$, simultaneously increase, it appears as if the productivity shock has given rise to an asset and a macroeconomic boom. But the productivity shock has had sectoral consequences; the rise in $\phi^*$ crowds out capital investment in the productive sector, therefore reducing $K_{s+1}$ and the growth rate of capital thereafter, compared to what it would have been had the shock not occurred. Accordingly, eventually output is lower than it otherwise would have been. The boom in the real estate sector has markedly different effects in the short term and in the long term.[32] We summarize this implication as the following Proposition.

**Proposition 7 The short- and long-term impact of a real estate boom**

*Although a productivity increase in the real estate sector from a higher value of* $\nu$ *generates temporary output (GDP), consumption, and land price booms in the short term, it leads to slower productivity and economic growth in the long term, thus lowering the wages and welfare of future generations.*[33]

Whether social welfare (as assessed, for instance, by a standard utilitarian social welfare function, discounting future generations utility) is increased or decreased depends on the discount rate. With sufficiently low discount rates, social welfare is decreased.[34]

**Discussion 3. Related literature**

Our paper is related to the vast modern literatures on credit and financial frictions, land, monetary economics, OLG, and endogenous growth. It brings together in a parsimonious model key insights from each, noting how, in many instances, standard results are overturned once these more constrained models are extended. A key lesson is that we need to be careful in drawing implications for monetary policy in models without land; without endogenous growth; without credit frictions, without a sensitivity to how monetary policy affects type of credit or sectoral credit expansion, i.e., whether credit expansion is based on land collateral financing or capital collateral financing.

---

[32] Even if the productivity increase is temporary, in our framework with endogenous growth, this leads to a decrease in capital investment and lowers the level of capital accumulation permanently, thereby leading to decreased wages and welfare of future generations.

[33] In the case of a low-interest-rate policy or an increase in land leverage, with those policy changes, output (GDP) remains unchanged, but aggregate consumption and land price booms are generated together with simultaneous increases in all ratios noted in Corollary.

[34] As Ramsey (1928) argued, there is no ethical justification for discounting future generations.

The modern macro-finance literature, including seminal papers by Stiglitz and Weiss (1981), Bernanke (1983), and Diamond (1984), emphasizes the role of credit and the restrictions in the provision of credit to entrepreneurs. By putting credit at the center stage of macroeconomic analysis, the literature has deepened our understanding of how credit-driven macroeconomic fluctuations occur (see Greenwald, Stiglitz, and Weiss 1984; Stiglitz and Weiss 1986, 1992; Woodford 1988; Bernanke and Gertler 1989; and Greenwald and Stiglitz 1993 and Stiglitz and Greenwald 2003 for earlier work). The focus of these papers was on total credit availability. In contrast, our paper emphasizes type of credit expansion or sectoral credit expansion.[35]

Based on the above fundamental theoretical papers, Kiyotaki and Moore (1997) advanced the literature by incorporating asset prices explicitly (see also Bernanke, Gertler, and Gilchrist 1999). In their paper and many of subsequent papers,[36] land plays an important role as collateral. An increase in the value of land relaxes borrowing constraints of entrepreneurs who have productive investment opportunities, allowing them to make more of their productive investments, increasing the efficiency of macroeconomy. But it is assumed by model construction that the only thing productive entrepreneurs can do with borrowed funds is to engage in entrepreneurial activity with high returns. In contrast, in our model, entrepreneurs decide an optimal portfolio allocation between different assets, while in most of the macro-finance literature, there is no such portfolio selection problem. Entrepreneurs may use borrowed funds for land speculation as well as for productive investment. An increase in the collateral value of land produces a crowding in effect, as in their model, but through a quite different mechanism, i.e., in our model, a lower downpayment to buy real estate leaves more funds available for capital investment. More importantly, unlike their model, the increase in the collateral value of land produces a general equilibrium crowding out effect, as entrepreneurs' portfolios shift toward land speculation, rather than productive investments, with the increased land speculation crowding out more productive capital investments. We provide the conditions under which the crowding out effect of land speculation outweighs the crowding in effect, suggesting a strong presumption that the former effect normally dominates, a result consistent with empirical results and common interpretations of many recent episodes of credit expansions (Verner 2019; Müller and Verner 2023; Banerjee, Mehrotra, and Zampolli 2024; Dai, Müller and Verner 2025).

[35] Matsuyama (2013) emphasizes credit composition and show that changes in the composition lead to endogenous macroeconomic fluctuations. While Matsuyama (2013) abstracts from asset prices, our focus is on the long-term impact of asset price booms with credit expansion. see also Greenwald and Stiglitz (1993), who show endogenous fluctuations.

[36] Representative papers on the financial accelerator includes Bernanke and Gertler (1989), Kiyotaki and Moore (1997), Bernanke, Gertler, and Gilchrist (1999), Krishnamurthy (2003), He and Krishnamurthy (2013), Brunnermeier and Sannikov (2014), Gârleanu and Panageas (2023), among others. These papers present a theory that explains asset prices, credit expansion, and macroeconomic virtuous cycles. Allen, Barlevy, and Gale (2022) is an exception. They show that during asset booms, borrowers undertake excessively risky investments and misallocation of resources occurs. A comprehensive literature review is too large for this paper; for a more complete discussion, see the 2022 Nobel Memorial Prize in Economics https://www.nobelprize.org/uploads/2022/10/advanced-economicsciencesprize2022-2.pdf

Throughout the present paper, we have emphasized that the macroeconomic consequences are markedly different between two-asset models of capital and money such as Tobin (1965) and our three-asset model including land. Brunnermeier and Sannikov (2016) have recently revive Tobin's intuition in an economy with incomplete markets, i.e., higher inflation due to higher money growth lowers the real interest rate on money and induces the portfolio choice towards capital investment. They conclude that moderate inflation boosts the growth rate of the economy and welfare, providing a micro-foundation for an inflation target. Their model abstracts from the existence of land. In contrast, in our model with land, to ensure that inflation targeting leads to an increased overall welfare in the long term, financial regulations on land related financing need also be introduced so that more funds flow into capital investment rather than land speculation.

Our paper is also related to the effect of land in overlapping generations models. There are two substantial differences of our paper from that literature. First, the welfare implications of the existence of land in our paper are markedly different from those in standard overlapping generations models with land. In the standard model with exogenous growth (including no growth), without land, the growth rate of the economy can be greater than the interest rate and there can be capital over-accumulation (Diamond (1965)), producing inefficient equilibria. The existence of land crowds out capital, but this is a good crowding-out effect; the existence of land eliminates inefficient equilibria, thereby increasing welfare (McCallum 1987; Mountford 2004). In contrast, in our model with *endogenous* growth, regardless of whether the growth rate is greater or lower than the interest rate, the very existence of land *may* reduce overall economic welfare by lowering rates of growth as real estate investment crowds out productive investment. Second, we study the effects of monetary policies and financial de-regulations on long-term productivity and economic growth, depending on the size of crowding-in and-out effects.

Finally, our paper is related to Benigno and Fornaro (2014) and Benigno, Fornaro, and Wolf (2025), who construct small open-economy models with tradable and non-tradable sectors, where the tradable sector is a growth-enhancing sector, and show that funds flowing into the non-tradable sector lowers long-term growth, a result totally consistent with ours (their non-growth enhancing sectors is the non-tradable sector, ours is real estate). There are three key differences. First, in our model, the type of credit expansion, whether it is land collateral financing, and asset prices play critical roles for whether policy changes are growth-enhancing, while they abstract from asset prices and credit friction; in their papers, a consumption boom associated with an expansion in the non-tradable sector drives funds to the non-growth sector. Second, in our model, monetary policy changes affect the real economy in the short run and long run through *endogenous* changes in the interest rates in the general equilibrium system, while in theirs, interest rates are exogenous. Third, as a difference in modeling methods, to illustrate the key idea, we employ a one-sector and closed-economy model, where land, combined with labor and capital, is a necessary input for production, while they employ the two-sector open economy models.

**5. Conclusion**

Based on the recent empirical findings on the macroeconomic impact of credit expansions, we have constructed a model that studies the long-run impact of credit expansion due to increases in land or capital collateral financing or a low-interest-rate policy.

There are three central messages of this paper. First, it is not so much aggregate credit expansion that matters for long-run productivity and economic growth but the type of credit expansions. Credit expansions associated mainly with relaxation of real estate financing (of capital investment financing) will be productivity-and growth-reducing (enhancing) in the long run. These results are consistent with empirical results of many recent episodes of credit expansions (Verner 2019; Müller and Verner 2023; Banerjee, Mehrotra, and Zampolli 2024; Dai, Müller and Verner 2025).

Second, lower interest rates and easier monetary policy may (as happened in the run-up to the 2008 financial crisis) induce more real estate speculation than productive capital investment; indeed the latter may be reduced, and with it economic growth, even though such policy changes generate temporary asset booms. These macroeconomic consequences are just the opposite of what was intended, and predicted by many of the standard models—including that of Tobin (1965)-- which ignored the presence of land as an alternative (unproductive) store of value. Hence, to ensure enhanced growth, looser monetary policy has to be accompanied by financial regulations that ensure that funds flow into the productive sectors of the economy.

Third, the welfare impact of asset price booms due to monetary easing policy differs substantially for agents of different generations and types, in particular, the generation that experienced the asset price boom and the subsequent generations who suffer from the subsequent lower growth. Although policies generating temporary asset price booms benefit the initial asset holder and borrowers, the lower interest rates and slower economic growth that they induce reduce welfare for future generations—both entrepreneurs and workers.

## Appendix

### A1. Deriving optimal portfolio selection

We construct the Lagrangian function,

(A2-1) $$\mathcal{L}_t^i = R^c k_{t+1}^i + R_t^x P_t x_t^i + Q_{t+1} m_t^i - (1+r_t)\left(k_{t+1}^i + P_t x_t^i + Q_t m_t^i - w_t\right)$$

$$+ \Lambda_{1,t}^i\left[\theta R^c k_{t+1}^i + \theta^x R_t^x P_t x_t^i - (1+r_t)\left(k_{t+1}^i + P_t x_t^i + Q_t m_t^i - w_t\right)\right] + \Lambda_{2,t}^i m_t^i$$

where $\Lambda_{1,t}^i \geq 0$ $\Lambda_{2,t}^i \geq 0$ are the Lagrangian multiplier associated with the borrowing and short-sale constraints.

The first-order necessary conditions are:

(A2-2) $$\frac{\partial \mathcal{L}_t^i}{\partial k_{t+1}^i} = R^c - (1+r_t) + \Lambda_{1,t}^i[\theta R^c - (1+r_t)] = 0$$

(A2-3) $$\frac{\partial \mathcal{L}_t^i}{\partial P_t x_t^i} = R_t^x - (1+r_t) + \Lambda_{1,t}^i[\theta^x R_t^x - (1+r_t)] = 0$$

(A2-4) $$\frac{\partial \mathcal{L}_t^i}{\partial Q_t m_t^i} = \frac{Q_{t+1}}{Q_t} - (1+r_t) - \Lambda_{1,t}^i(1+r_t) + \frac{\Lambda_{2,t}^i}{Q_t} = 0$$

Complementary slackness conditions:

(A2-5a) $\Lambda_{1,t}^{i}\left[\theta R^{c}k_{t+1}^{i}+\theta^{x}R_{t}^{x}P_{t}x_{t}^{i}-(1+r_{t})(k_{t+1}^{i}+P_{t}x_{t}^{i}-w_{t})\right]=0$ and $\Lambda_{t}^{i}\geq 0$.

(A2-5b) $\Lambda_{2,t}^{i}m_{t}^{i}\geq 0$ and $m_{t}^{i}\geq 0$.

By rearranging (A2-2), we have

(A2-6) $\Lambda_{1,t}^{i}=\frac{R^{c}-(1+r_{t})}{1+r_{t}-\theta R^{c}}$

Similarly, from (A2-3), we have

(A2-7) $\Lambda_{1,t}^{i}=\frac{R_{t}^{x}-(1+r_{t})}{1+r_{t}-\theta^{x}R_{t}^{x}}$

It follows that if and only if $R^{c}>1+r_{t}$ and $R_{t}^{x}>1+r_{t}$, $\Lambda_{1,t}^{i}>0$, i.e., the borrowing constraint binds, provided that $1+r_{t}>\theta R^{c}$ and $1+r_{t}>\theta^{x}R_{t}^{x}$, which is ensured by Assumption 1. Also, in equilibrium, (2) holds. Then, from (A2-4), we have $\Lambda_{1,t}^{i}(1+r_{t})=\frac{\Lambda_{2,t}^{i}}{Q_{t}}$. Therefore, when $\Lambda_{1,t}^{i}>0$, $\Lambda_{2,t}^{i}>0$, i.e., the short-sale constraint binds.

From (A2-6) and (A2-7),

(A2-8) $\frac{R^{c}-(1+r_{t})}{1+r_{t}-\theta R^{c}}=\frac{R_{t}^{x}-(1+r_{t})}{1+r_{t}-\theta^{x}R_{t}^{x}},$

which can be written as

(A2-9) $\frac{R^{c}(1-\theta)-(1+r_{t})+\theta R^{c}}{1+r_{t}-\theta R^{c}}=\frac{R_{t}^{x}(1-\theta^{x})-(1+r_{t})+\theta^{x}R_{t}^{x}}{1+r_{t}-\theta^{x}R_{t}^{x}},$

which is equivalent to (7b),

**A2. The existence of a unique balanced growth path**

At the balanced growth path, we know that

(16) $1 + r^* = \frac{1+g^*}{1+\mu}$.

(12b) $1 + g^* = \frac{\lambda}{1-\theta^x+\frac{\theta^x\lambda}{1+r^*}} / (1 + \epsilon/\phi^*)$.

where, it will be recalled, $\lambda \equiv \frac{R^C(1-\theta)}{1-\frac{\theta R^C}{1+r^*}}$. It follows from straightforward substitution that

(18a) $1 + r^* = \frac{R^C}{1-\theta^x}\left[\frac{1-\theta}{(1+\mu)(1+\epsilon/\phi^*)} - (\theta^x - \theta)\right]$.

Using (18a), we have

(A4-1) $1 - \theta^x + \frac{\theta^x\lambda}{1+r^*} = \frac{1-\theta^x}{1-\theta^x(1+\mu)(1+\epsilon/\phi^*)}$.

Substituting (A4-1) into (12b) yields the quadratic equations regarding $\phi^*$.

(A4-2) $A[1-\theta^x(1+\mu)](\phi^*)^2 + \Big\{-\eta\rho A - A\theta^x(1+\mu)\epsilon + A\epsilon[1-\theta^x(1+\mu)] +$

$\frac{R^C(1-\theta)}{1-\theta^x}[1-\theta^x(1+\mu)]\Big\}\phi^* - \epsilon\Big\{\eta\rho A + A\theta^x(1+\mu)\epsilon + \frac{R^C(1-\theta)}{1-\theta^x}\theta^x(1+\mu)\Big\} = 0.$

Since the left-hand side of (A4-2) is a convex function of $\phi^*$, with the negative value at $\phi^* = 0$ if $\epsilon > 0$, it is clear from (A4-2) that, one of the solutions is positive and the other one is negative. When $\epsilon = 0$, the other solution is zero. Hence, for $\epsilon \geq 0$, there exists a unique balanced growth path in which $\phi^* > 0$ is constant. It is obvious from (18a) and (12b) that when $\phi^* > 0$ is constant, $1 + r^*$ and $1 + g^*$ are constant and are uniquely determined.

### A3. Dynamic stability

As noted in the main text, the dynamics of this economy can be characterized by (2), (9), (10), (12b), (14a) and (15b).

Using (7b), (12b) can be written as

(A5-1) $\phi_{t+1} = \left( \frac{\frac{\lambda_t}{1-\theta^x + \frac{\theta^x \lambda_t}{1+r_t}}}{\frac{1}{1-\frac{\theta R^c}{1+r_t}} \left[ \eta\rho - \left( \frac{1-\theta^x}{1-\theta^x + \frac{\theta^x \lambda_t}{1+r_t}} \right) \phi_t \right] A} \right) \phi_t - \epsilon = \psi(\phi_t, 1+r_t).$

where $\lambda_t \equiv \frac{R^c(1-\theta)}{1-\frac{\theta R^c}{1+r_t}}$.

Using (9), (14a), and (15b), (2) can be written as

(A5-2) $1 + r_t = \frac{Q_{t+1}}{Q_t} = \frac{1}{1+\mu} \left[ \frac{\left\{ \rho - \frac{1}{1-\frac{\theta R^c}{1+r_{t+1}}} \left[ \eta\rho - \left( \frac{1-\theta^x}{1-\theta^x + \frac{\theta^x \lambda_{t+1}}{1+r_{t+1}}} \right) \phi_{t+1} \right] - \phi_{t+1} \right\}}{\left\{ \rho - \frac{1}{1-\frac{\theta R^c}{1+r_t}} \left[ \eta\rho - \left( \frac{1-\theta^x}{1-\theta^x + \frac{\theta^x \lambda_t}{1+r_t}} \right) \phi_t \right] - \phi_t \right\}} \right] \left\{ \frac{1}{1-\frac{\theta R^c}{1+r_t}} \left[ \eta\rho - \right. \right.$

$\left. \left. \left( \frac{1-\theta^x}{1-\theta^x + \frac{\theta^x \lambda_t}{1+r_t}} \right) \phi_t \right] A \right\},$

giving $r_{t+1}$ as a function of $\{r_t, \phi_t, \phi_{t+1}\}$. Using (A5-1), we have

(A5-3) $1 + r_{t+1} = \omega(\phi_t, 1 + r_t).$

Hence, the dynamics of the monetary economy can be characterized by (A5-1) and (A5-3), so it is reduced to the two-dimensional dynamical system, where both $\phi_t$ and $1 + r_t$ are jump variables.

We know from the analysis in Appendix A2 that there exists a unique balanced growth path (steady state). Hence, there exists a rational expectation path along which the economy instantly achieves the balanced growth path.

**A4. Proof of Proposition 7**

By differentiating (18b) and (18c), we obtain

$$\lim_{\epsilon\to 0}\frac{d\phi^*}{d\epsilon}=\frac{\frac{R^c(1-\theta)A[1-\theta^x(1+\mu)]^2}{1-\theta^x}+A\theta^x(1+\mu)\eta\rho}{[1-\theta^x(1+\mu)]\left\{\eta\rho A-\frac{R^c(1-\theta)[1-\theta^x(1+\mu)]}{1-\theta^x}\right\}}>0.$$

$$\lim_{\epsilon\to 0}\frac{d(1+g^*)}{d\epsilon}=-\frac{R^c(1-\theta)}{(1-\theta^x)\phi^*}<0.$$

**Supplementary material for "Credit Expansion, Land Speculation, and Low-Interest-Rate Policy"** by Tomohiro Hirano and Joseph E. Stiglitz

In the main text, we presented a bare-bones model integrating credit frictions and endogenous growth. For tractability, we have made a number of simplifying assumptions concerning preferences and technology. For instance, the real estate sector uses land only, agents have linear utility over consumption only when old, and workers cannot invest in real estate. In this on-line appendix, we show that the results can be generalized.

**O1. Robustness:** ***Weakening the assumptions of limited participation in the credit market and credit frictions***

We need to distinguish between the return on a safe asset (a government bond. If there is uncertainty about the rate of inflation, bonds that are not inflation protected are still risky.) and a return on a risky investment. Both land and capital are risky, and not surprisingly, on average yield substantially higher returns than government bonds. This is what we would expect in equilibrium. Credit frictions inevitably follow, not only from asymmetries of information, but also from limited liability (or more generally, even in the absence of limited liability, the limited ability to recover funds.)

Think, for a moment, of workers as infinitely risk averse. Then they would only be willing to lend a limited amount either to an entrepreneur buying capital or buying land. There is a natural limit on leverage, even in the absence of policy. Those limits on leverage will, of course, mean that even with diminishing returns in production, the average return on capital and on land will remain above the safe rate of interest. Similar results would hold in a model with highly risk averse workers.

The model we have formulated here captures in a simple way these credit frictions, though crucially, we have not explicitly modelled either asymmetries of information or risk aversion. But it explains clearly why, for instance, there can be an equilibrium with productive land having a finite price, even if $r$ is very low. The non-entrepreneurial savers see (rightly) land holdings as risky, and so even if $r$ were zero, they would only have a finite demand for it, and its price would be finite.

Similarly, if government were to issue an excessive number of bonds, capital accumulation would be limited, the return on capital would increase, and the private sector could issue an increasing number of safe bonds: there is thus a limit to the *equilibrium* issuance of bonds by government at low interest rates at an interest rate less than $g$. Thus, while it might *seem* that with low $r$, government debt *could* be unbounded, the interest rate is low only because government borrowing is limited.

Thus, what have been posed as critical conundrums in the literature are really a reflection of the oversimplifications employed in the models in which they arise and the attempt to interpret the real world's data through the lens of those models.

**O2. Robustness: Analysis of a different borrowing constraint**

For the robustness of our main results, we study the case with a different borrowing constraint.

(O3-1) $b_t^i \leq \theta k_{t+1}^i + \theta^x P_t x_t^i.$

This collateral constraint is set as a fraction of this year's value of land.

With this borrowing constraint, (7a) changes to

$$\text{(O3-2)} \qquad \underbrace{\frac{R_t^c-(1+r_t)\theta}{1-\theta}}_{\substack{\text{leveraged rate of return}\\ \text{on capital investment}}} = \underbrace{\frac{R_t^x-(1+r_t)\theta^x}{1-\theta^x}}_{\substack{\text{leveraged rate of return}\\ \text{on land speculation}}}.$$

Solving for $R_t^x$ yields

$$\text{(O3-3)} \quad R_t^x = \left(\frac{R_t^c-(1+r_t)\theta}{1-\theta}\right)(1-\theta^x) + (1+r_t)\theta^x = \frac{R_t^c(1-\theta^x)+(1+r_t)(\theta^x-\theta)}{1-\theta}.$$

The capital-investment function of entrepreneurs (8) changes to

$$\text{(O3-4)} \quad k_{t+1}^i = \underbrace{\frac{1}{1-\theta}}_{\text{leverage}} \left[ \underbrace{w_t}_{\text{saving}} - \underbrace{(1-\theta^x)P_t x_t^i}_{\substack{\text{total down-payment}\\ \text{in buying land}}} \right].$$

The dynamics equations of (11), (14a), (12b), and (15b) change to

$$\text{(O3-5)} \quad K_{t+1} = \frac{1}{1-\theta}[\eta w_t - (1-\theta^x)P_t] = \frac{1}{1-\theta}[\eta(1-\alpha)AK_t - (1-\theta^x)P_t].$$

$$\text{(O3-6)} \quad 1 + g_t \equiv \frac{K_{t+1}}{K_t} = \frac{1}{1-\theta}[\eta(1-\alpha) - (1-\theta^x)\phi_t]A.$$

$$\text{(O3-7)} \quad \phi_{t+1} = \phi_t R_t^x / \left\{\frac{1}{1-\theta}[\eta(1-\alpha) - (1-\theta^x)\phi_t]A\right\} - \epsilon a^\alpha.$$

$$\text{(O3-8)} \quad Q_t M_t = \left\{(1-\alpha) - \frac{1}{1-\theta}[\eta(1-\alpha) - (1-\theta^x)\phi_t] - \phi_t\right\} AK_t,$$

The dynamics of this economy, entailing $\{P_t, K_t, Q_t, M_t, r_t, T_t\}$ can be characterized by (2), (9), (10), (O3-6), (O3-7), and (O3-8).

First, we proceed with our analysis by focusing on steady states.

***Steady states: unproductive land*, i.e., $\epsilon = 0$**

When $\epsilon = 0$, we obtain

(O3-9) $1 + r^* = \frac{R^c(1-\theta^x)}{1-\theta^x+\mu(1-\theta)}$.

(O3-10) $\phi^* = \frac{\eta(1-\alpha)}{1-\theta^x} - \frac{R^c(1-\theta^x)(1+\mu)}{A[1-\theta^x+\mu(1-\theta)]}$.

(O3-11) $1 + g^* = \frac{R^c(1-\theta^x)(1+\mu)}{1-\theta^x+\mu(1-\theta)}$.

**General case**

In steady state, the real value of money grows according to

(16) $1 + r^* = \frac{Q_{t+1}}{Q_t} = \frac{1+g^*}{1+\mu}$.

(16), (O3-6), and (O3-7) give three equations in three unknowns, $r^*$, $g^*$, and $\phi^*$. By rearranging them, we can derive the quadratic equation for $\phi^*$.

(O3-12) $\frac{A(1-\theta^x)H}{1-\theta}(\phi^*)^2 + \left[-\frac{A\eta s(1-\alpha)H}{1-\theta} + R^c(1-\theta^x)(1+\mu) + (1-\theta^x)(1+\mu)\epsilon a^\alpha\right]\phi^* - A\eta(1-\alpha)(1+\mu)\epsilon a^\alpha = 0$,

where $H = 1 - \theta^x + \mu(1-\theta)$. Since the left-hand side of (O3-9) is a convex function of $\phi^*$, with the negative intercept, there exists a unique value of $\phi^* > 0$, while the other solution is negative (when $\epsilon = 0$, the other solution is $\phi^* = 0$). Solving for the positive $\phi^* > 0$ yields

(O3-13) $\phi^* = \frac{-\Upsilon + \sqrt{\Upsilon^2 + 4\frac{A(1-\theta^x)H}{1-\theta}A\eta(1-\alpha)(1+\mu)\epsilon a^\alpha}}{2\frac{A(1-\theta^x)H}{1-\theta}}$,

where $\Upsilon = -\frac{A\eta(1-\alpha)H}{1-\theta} + R^c(1-\theta^x)(1+\mu) + (1-\theta^x)(1+\mu)\epsilon a^\alpha$.

By substituting (O3-13) into (O3-6), we obtain the steady-state growth rate of aggregate capital.

$$\text{(O3-14)}\quad 1+g^* = \frac{A}{1-\theta}\left[\eta(1-\alpha)-(1-\theta^x)\frac{-\Upsilon+\sqrt{\Upsilon^2+4\frac{A(1-\theta^x)H}{1-\theta}A\eta(1-\alpha)(1+\mu)\epsilon a^\alpha}}{2\frac{A(1-\theta^x)H}{1-\theta}}\right].$$

The steady-state interest rate is given by

$$\text{(O3-15)}\quad 1+r^* = \frac{A}{(1+\mu)(1-\theta)}\left[\eta(1-\alpha)-(1-\theta^x)\frac{-\Upsilon+\sqrt{\Upsilon^2+4\frac{A(1-\theta^x)H}{1-\theta}A\eta(1-\alpha)(1+\mu)\epsilon a^\alpha}}{2\frac{A(1-\theta^x)H}{1-\theta}}\right].$$

For $\phi^*$ to be positive, even when $\epsilon = 0$, we make the following assumption.

**Assumption 3.** $\frac{\eta(1-\alpha)A}{1-\theta^x} > \frac{R^c(1-\theta)(1+\mu)}{1-\theta^x+\mu(1-\theta)}$.

Also, for $Q_t$ to be positive, we impose conditions on the parameter values so that the following inequality condition must be satisfied.

**Assumption 4.** $A(1-\alpha) - \frac{A}{1-\theta}[\eta(1-\alpha)-(1-\theta^x)\phi^*] - \phi^* = A(1-\alpha)-(1+g^*) - A\phi^* > 0$, where $1+g^*$ and $\phi^*$ are given by (O3-14) and (O3-15).

For instance, when $\epsilon \to 0$, Assumption 4 can be written as

**Assumption 4'.** $A(1-\alpha) > \frac{\eta(1-\alpha)A}{1-\theta^x}$.

Combining Assumptions 3 and 4' leads to

$$A(1-\alpha) > \frac{\eta(1-\alpha)A}{1-\theta^x} > \frac{R^c(1-\theta)(1+\mu)}{1-\theta^x+\mu(1-\theta)}.$$

Then, $A(1-\alpha) > \frac{R^c(1-\theta)(1+\mu)}{1-\theta^x+\mu(1-\theta)}$ can be written as

$$(1-\alpha)[1-\theta^x+\mu(1-\theta)] > \left(\alpha+\frac{1-\delta}{A}\right)(1-\theta)(1+\mu),$$

which is likely to be satisfied as $\delta$ and/or $A$ are large enough, and $\alpha$ is small enough. Therefore, by choosing $\eta$ appropriately, we can verify that there are relevant parameters for which Assumptions 3 and 4' are simultaneously satisfied.

Then, we obtain the following Proposition.

**Proposition A2 Existence of balanced growth path**

Under Assumption 3 and 4, for $\epsilon \geq 0$, there exists a unique balanced growth path where the relative size of land speculation $\phi^* > 0$, economic growth rate $1 + g^*$, the interest rate $1 + r^*$, which equals the rate of return on fiat money $\frac{Q_{t+1}}{Q_t}$, are constant over time. $\phi^*$, $1 + g^*$, and $1 + r^*$ are given by (O3-13), (O3-14), and (O3-15), respectively.

**Comparative statics**

(O3-13), (O3-14), and (O3-15) allow us to conduct comparative statics assessing the effects of changes in the market/policy parameters, with results and intuitions analogous to the ones in the main text. We can obtain the following results concerning comparative statics and dynamic stability.

**Proposition A3 Impact of changes in the collateral values or in monetary policy**

Under Assumption 3 and 4, for $\epsilon \geq 0$, we have

(A3-i) $\frac{d(1+g^*)}{d\theta^x} < 0$, with $\frac{d\phi^*}{d\theta^x} > 0$ and $\frac{d(1+r^*)}{d\theta^x} < 0$.

(A3-ii) $sign \frac{d(1+g^*)}{d\mu} = sign(\theta - \theta^x)$, with $sign \frac{d\phi^*}{d\mu} = sign(\theta^x - \theta)$ and $\frac{d(1+r^*)}{d\mu} < 0$.

(A3-iii) $\frac{d(1+g^*)}{d\theta} > 0$, with $\frac{d\phi^*}{d\theta} > 0$ and $\frac{d(1+r^*)}{d\theta} > 0$.

**Proof**:

Since direct calculations using (O3-13), (O3-14), and (O3-15) are complicated, we will take another route for comparative statics. Since $1 + g^*$ is determined according to (O3-6), we will examine how $(1 - \theta^x)\phi^*$ is affected by changes in $\{\mu, \theta^x, \theta\}$.

(O3-12) is equivalent to

$$\Omega((1-\theta^x)\phi^*) \equiv \frac{AH}{(1-\theta)(1-\theta^x)(1+\mu)}\big((1-\theta^x)\phi^*\big)^2 + \left[-\frac{A\eta s(1-\alpha)H}{(1-\theta)(1-\theta^x)(1+\mu)} + R^c + \epsilon a^\alpha\right](1-\theta^x)\phi^* - A\eta(1-\alpha)\epsilon a^\alpha = 0.$$

Since $\Omega$ is a convex function of $(1-\theta^x)\phi^*$, with the negative intercept, there exists a unique value of $(1-\theta^x)\phi^* > 0$, while the other solution is negative, i.e., $\phi^* < 0$. In addition, for any value of $(1-\theta^x)\phi^* < A\eta(1-\alpha)$(otherwise $1 + g^*$ would be negative in (O3-6), the following results are obtained.

First, concerning (A3-i), we have

$$\frac{d\Omega}{d\theta^x} = \frac{AH\mu(1-\theta)}{(1-\theta)(1+\mu)(1-\theta^x)^2}\big((1-\theta^x)\phi^* - A\eta(1-\alpha)\big)(1-\theta^x)\phi^* < 0.$$

This means $\frac{d(1-\theta^x)\phi^*}{d\theta^x} > 0$ (implies $\frac{d\phi^*}{d\theta^x} > 0$). Therefore, $\frac{d(1+g^*)}{d\theta^x} < 0$.

Second, concerning (A3-ii), we have

$$\frac{d\Omega}{d\mu} = \frac{AH(\theta^x-\theta)}{(1-\theta)(1-\theta^x)(1+\mu)^2}\big((1-\theta^x)\phi^* - A\eta(1-\alpha)\big)(1-\theta^x)\phi^*.$$

This means $sgn\frac{d(1-\theta^x)\phi^*}{d\mu} = sgn(\theta^x - \theta)$. Therefore, $sign\frac{d(1+g^*)}{d\mu} = sign(\theta - \theta^x)$.

For $\frac{d(1+r^*)}{d\mu} < 0$, in the steady-state equilibrium, we know

$$1 + r^* = \frac{1+g^*}{1+\mu} = \frac{R^c(1-\theta^x)\phi^*}{[(1+\mu)(1-\theta)-(\theta^x-\theta)]\phi^* + (1+\mu)(1-\theta)\epsilon a^\alpha}.$$

When $\theta^x < \theta$, $\frac{d\phi^*}{d\mu} < 0$. Therefore, the right hand side of the above equation decreases as $\mu$ increases. On the other hand, when $\theta^x > \theta$, $\frac{d(1+g^*)}{d\mu} < 0$. Therefore, it is obvious that the interest rate decreases.

Finally, concerning (A3-iii), we have

$$\frac{d\Omega}{d\theta} = \frac{AH(1-\theta^x)}{(1+\mu)(1-\theta^x)(1-\theta)^2}\big((1-\theta^x)\phi^* - A\eta(1-\alpha)\big)(1-\theta^x)\phi^* < 0.$$

This means $\frac{d(1-\theta^x)\phi^*}{d\theta} > 0$, i.e., $\frac{d\phi^*}{d\theta} > 0$.

Also,

$$1 + g^* = \frac{R^c(1-\theta^x)(1+\mu)\phi^*}{[(1+\mu)(1-\theta)-(\theta^x-\theta)]\phi^* + (1+\mu)(1-\theta)\epsilon a^\alpha}.$$

Since we know that $\frac{d\phi^*}{d\theta} > 0$, the right-hand side of the above equation is an increasing function of $\theta$. Therefore, $\frac{d(1+g^*)}{d\theta} > 0$.

**Dynamic stability**

We can analyse dynamic stability. The following Proposition summarizes it.

**Proposition A4 Uniqueness of Rational Expectations Trajectory**

There exists a unique rational expectations path along which the economy achieves the balanced growth path immediately without transitional dynamics. The economy immediately converges to that path, through the setting of the initial price of land, $P_0 = AK_0\phi^*$.

**Proof**

Using (O3-3), (O3-7) can be written as

$$\phi_{t+1} = \left( \frac{\frac{R^c(1-\theta^x)+(1+r_t)(\theta^x-\theta)}{1-\theta}}{\frac{1}{1-\theta}[\eta(1-\alpha)-(1-\theta^x)\phi_t]A} \right) \phi_t - \epsilon a^\alpha = \psi(\phi_t, 1+r_t).$$

Using (O3-6) and (O3-8), (16) can be written as

$$1 + r_t = \frac{Q_{t+1}}{Q_t} = \frac{1}{1+\mu} \left[ \frac{\left\{(1-\alpha)+(1-\eta)e(a)^\alpha - \frac{1}{1-\theta}[\eta(1-\alpha)-(1-\theta^x)\phi_{t+1}]-\phi_{t+1}\right\}}{\left\{(1-\alpha)+(1-\eta)e(a)^\alpha - \frac{1}{1-\theta}[\eta(1-\alpha)-(1-\theta^x)\phi_t]-\phi_t\right\}} \right] \left\{ \frac{1}{1-\theta} [\eta(1-\alpha) - (1 - \theta^x)\phi_t]A \right\} = \omega(\phi_t, \phi_{t+1}),$$

giving $1 + r_t$ as a function of $\{\phi_t, \phi_{t+1}\}$.

Hence, by substituting the relation $1 + r_t = \omega(\phi_t, \phi_{t+1})$ into $\phi_{t+1} = \psi(\phi_t, 1 + r_t)$, the dynamical system is reduced to the one-dimensional dynamical system for $\phi_t$. From Proposition A2, we know that for $\epsilon > 0$, there exists a unique steady-state value of $\phi^* > 0$ (the other steady-state value is negative), and $\phi_{t+1} = -\epsilon a^\alpha < 0$ when $\phi_t = 0$, and $\phi_t$ is a jump variable. Therefore, the only possible rational expectations trajectory is one that immediately achieves the steady state without transitional dynamics.

**O3. Robustness: More general production functions for real estate sector**

In this appendix, we extend the bare-bones model to the situation where both labor and land are inputs into real estate.

In the real estate sector, competitive firms produce the final consumption goods by using labor and land as factors for production. The aggregate production function is given by

(O3-1) $Y_t = D_t(N_t^X)^\rho (X)^{1-\rho}$,

where $N_t^X$ is labor input in the real estate sector.

The wage rates in the real estate sector and the productive sector are, respectively, given by

(O3-2) $w_t^X = \rho D_t (N_t^X)^{\rho-1}(X)^{1-\rho}$,

(O3-3) $w_t^K = (1-\alpha)A(N_t^K)^{-\alpha}K_t$,

where $N_t^K$ is labor input in the productive sector.

The labor market clearing condition is

(O3-4) $N_t^X + N_t^K = N = 1$.

We first consider a case where labor mobility is possible between the two sectors. Then, in equilibrium, the wage rates must be equal, i.e., $w_t^X = w_t^K \equiv w_t$. Considering $D_t = \epsilon a K_t$ and $X = 1$, we have

(O3-5) $(1-\alpha)A(N_t^K)^{-\alpha}K_t = \rho\epsilon a K_t (N_t^X)^{\rho-1}$,

which can be written as

(O3-6) $(1-\alpha)A(1-N_t^X)^{-\alpha} = \rho\epsilon a (N_t^X)^{\rho-1}$.

The left-hand side is a convex function of $N_t^X$ with a positive slope and a positive intercept $(1-\alpha)A$ at $N_t^X = 0$ and goes to infinity as $N_t^X$ goes to 1, while the right-hand side is also a convex function of $N_t^X$ with a negative slope and goes to infinity as $N_t^X$ goes to zero and becomes equal to $\rho\epsilon a$ at $N_t^X = 1$. Hence, $N_t^X$ is uniquely determined and is constant, and so is $N_t^K$.

Hence, total incomes of entrepreneurs from the two sectors can be written as

(O3-7) $\eta w_t^X N^X + \eta w_t^K N^K = \eta w_t = \eta(1-\alpha)A(N^K)^{-\alpha} K_t.$

Next, we examine a case where labor mobility is not possible. In this case, total wage incomes can be written as

(O3-8) $\eta w_t^X N^X + \eta w_t^K N^K = \eta[\rho\epsilon a(N^X)^{\rho-1} + (1-\alpha)A(N^K)^{-\alpha}]K_t.$

In (O3-7) and (O3-8), total incomes of entrepreneurs are a linear function of aggregate capital stock $K_t$. Hence, the analysis in the main text will apply by replacing $\eta w_t$ in (11) with (O3-7) or (O3-8).